\documentclass{article}

\usepackage[english]{babel}
\usepackage{tablefootnote}
\usepackage[letterpaper,top=2cm,bottom=2cm,left=3cm,right=3cm,marginparwidth=1.75cm]{geometry}

\usepackage{amsmath}
\usepackage{amssymb}
\usepackage{graphicx}
\usepackage[colorlinks=true, allcolors=blue]{hyperref}
\usepackage{xcolor}
\usepackage{natbib}
\usepackage{aas_macros}

\title{\vspace{-1.0cm}{\Large Next Generation Very Large Array Memo \#122} \\ Characterization of the synthesized beam with and without MID antennas in Mexico}
\author{Alfonso Trejo-Cruz, Roberto Galván-Madrid,  
\and Carlos Carrasco-González, Eric F. Jiménez-Andrade, Stan Kurtz, 
\and Jes\'us M. J\'aquez-Dom\'inguez, Alice Pasetto, Luis A. Zapata \\({\small Instituto de Radioastronom\'ia y Astrof\'isica, UNAM)}}
\begin{document}
\date{July 2, 2024}
\maketitle

\begin{abstract}

Synthesized beam (PSF) synthetic observations with and without the antennas in Mexico are analyzed. For a simple continuum observing setup, we generated visibility files and their associated PSF images for a grid of parameters (robust weighting, tapering, and declination). The tests were done for both the MID and MID+Spiral+Core configurations and their cropped versions without antennas in Mexico. We show that the performance of the Array, in terms of the beam properties, is in general significantly better when both the MID array antennas are present in Northern Mexico and observations target southern sources. 
At a declination of --40 deg, there are increments in the ellipticity of at least $\sim$ 1.3$\times$ and 1.2$\times$ for a tapering of 3.0 and 4.0 mas, if the antennas in Mexico are not included. 
For the parameter space tested, the changes in ellipticity of the MID and MID+Spiral+Core configurations differ by $\sim10$\%. 
Larger tapering values help to reduce the ellipticity for cropped configurations at all declinations, but it will impose more constraints in terms of angular resolution.
\end{abstract}

\section{Context and goals}

The current MID configuration is denoted by 28MOD \citep{Walker_102} and has seven antennas located in Northern Mexico. This effectively extends the North-South baselines of the array, resulting in a more circular synthesized beam (PSF) when observing southern sources, in particular for declination Dec $\leq~ \sim -20$ deg.
Therefore, those antennas are important in the overall context of the ngVLA, as the community is expected to actively pursue observations in that region of the sky and exploit synergies between the ngVLA and SKA.

In this memo, we explore the implications for the synthesized beam if the antennas in Mexico were not to be deployed. This would affect not only the performance of the MID configuration but also observing modes using the entire MID + Spiral + Core. We focus on characterizing the PSF using a continuum observing mode, where the configuration, declination, and the imaging weighting and tapering values are part of the parameter space.
The consequences of (not) having MID antennas in Northern Mexico for specific science cases will be explored in the future.

\section{Setup for synthetic observations and mapping}
\label{sec:setup}
In order to characterize the point spread function (PSF) or synthesized beam, we set a parameter space for both the creation of visibilities (\textit{uv} plane) and imaging stages. Visibility (MS) files were created using the \texttt{CASA} \citep{CASAteam_2022} simulator toolkit (\texttt {sm}) and the PSF images with the \texttt {tclean} task. Our parameter space uses $ -40 \leq \mathrm{Dec} \leq 45$, covered in 12 steps, every 15 and 5 deg for Dec $>$0 and $<$0, respectively, in order to better sample the response of the array for more challenging observing directions. For each of these the following parameters were used: an on-source time of 2 hrs resulting from the hour angle (HA) --1 to +1 hr, a total bandwidth of 2.0 GHz in a single channel, and a time averaging per visibility of 60 seconds. 
The synthetic observations presented here are only meant to make a basic characterization of the PSF.
Multiple observing blocks and calibration loops are not used, as we seek to characterize the PSF in relatively short observations.

For the imaging stage, we use robust values from R = --2.0 to --0.2 and from +1.2 to +2.0, for the MID+Spiral+Core and MID-only configurations, respectively. Steps of 0.2, in the Briggs weighting scheme were employed for both configurations. All images have an image size of 8192 pixels and a cell size of $\sim$ 0.027 mas. This corresponds to 1/20 of the angular resolution of the MID+Spiral+Core longest baseline and is chosen to over sample the synthesized beam core. 8192 pixels correspond to maps of 221 mas in size, adequate to properly grid visibilities from the Core array baselines.

The specific robust range per configuration, instead of choosing the whole range from --2.0 to +2.0, is intended to obtain more gaussian PSFs, and therefore to enforce the CASA synthesized beams to be representative of those. The wide skirts in the MID+Spiral+Core images are reduced when choosing the more negative robust values; the PSFs from MID-only images are of better quality when choosing parameters closer to natural weighting. Tapering in the imaging stage was added to mitigate the non-gaussian features (narrow core plus wide skirt) in the PSF, which is more relevant for the MID+Spiral+Core images. A detailed discussion of these properties is in \citet{rosero_55,rosero_65}. The list of tapering values used are 2, 3, 4, and 5 mas. While we tested smaller tapering values, e.g., 1.0 and 0.5 mas, the resulting PSF images still present the known central sharp spike combined with a wide skirt, for some of the robust/declination values. A PSF was produced for each of the parameter combinations, resulting in a total of 240 and 480 images, for the MID-only and MID+Spiral+Core configurations. 
In this memo we center in the cases more difficult to observe at southern declinations.
All data were produced with version  6.5.4 of \texttt {CASA}.

\subsection{Configurations}
\label{sec:configs}
\begin{figure}[!htb]
  \begin{minipage}[c]{0.6\textwidth}
    \includegraphics[width=\textwidth]{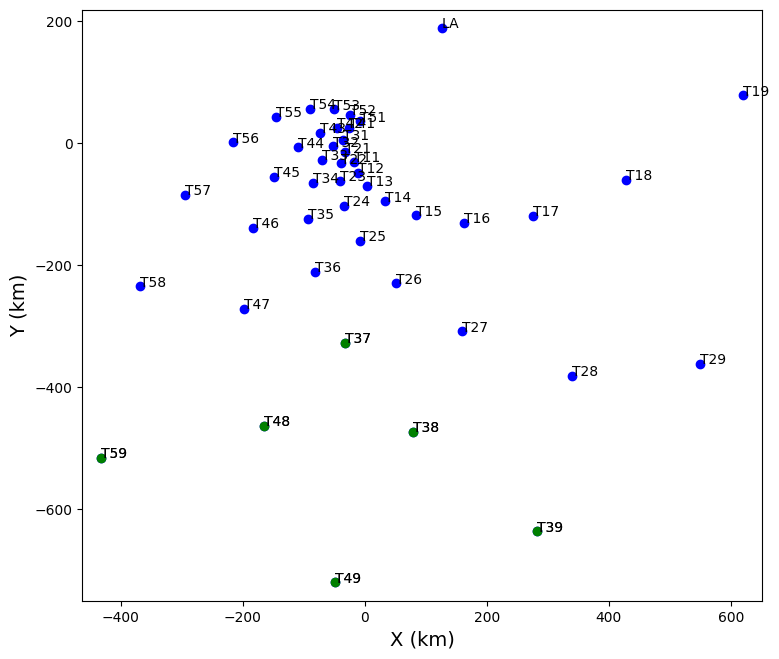}
  \end{minipage}\hfill
  \begin{minipage}[c]{0.35\textwidth}
    \caption{
       Distribution of MID (28MOD) antenna locations (blue plus green circles). The cropped MID configuration (without antennas in Mexico) is formed with the blue circles. The six antennas in Mexico are T37, T38, T39, T48, T49, and T59 and are shown in green circles. T27, which is about half a kilometer from the US-Mexico border, is not removed from the cropped configurations. 
       } \label{fig:locs}
  \end{minipage}
\end{figure}

Both the MID and MID+Spiral+Core configurations were used for the simulations, and for each of them we created an alternative configuration by removing the MID antennas located in Mexico (henceforth called the \textit{cropped} configurations). We kept T27 as part of these cropped configurations, even though that antenna is located in Mexico in 28MOD. T27 is located about half a kilometer from the US-Mexico border, next to the Río Bravo. Therefore, the results of our simulations will not change 
in a significant way if this antenna were to be relocated to the US, just across the border. 
The MID configuration used (28MOD), along with its cropped version, is presented in Fig. \ref{fig:locs}. 
Note that in the following sections we only discuss results for the case of ngVLA Band 6 centered at 93 GHz. 
The main results from this memo would also apply to the other bands at lower observing frequencies, when pixel sizes and tapering values are chosen appropriately. 

\begin{figure}[!htb]
\centering
\includegraphics[width=0.49\textwidth]{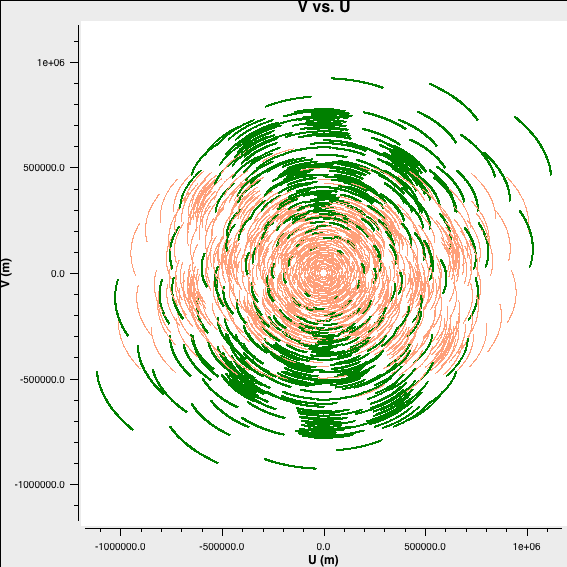}
\includegraphics[width=0.49\textwidth]{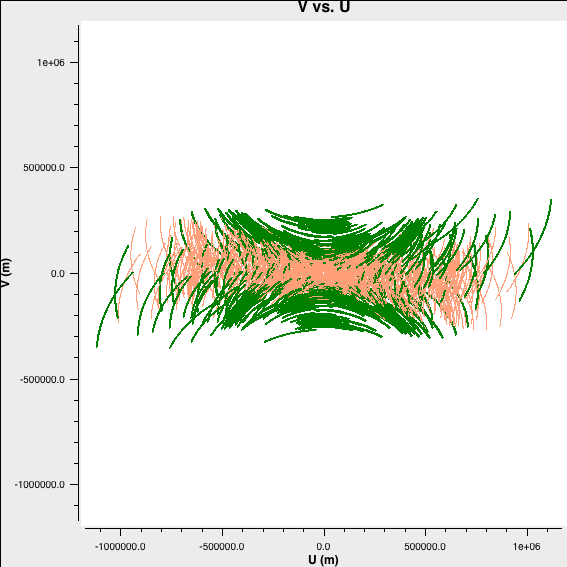}
\caption{\label{fig:uv_plane} Corresponding $uv$ coverage for the MID configurations shown in Fig. \ref{fig:locs}. The panels show ngVLA Band 6 cases centered at 93 GHz for Dec of +30 deg (\textit{left}) and --40 deg (\textit{right}). The \textit{cropped} array without antennas in Mexico produce the salmon data points; full configuration responses are formed by adding the green data points. Note the additional gaps and loss of mid-baselines in the North-South ($v$) direction resulting from the removal of the green data points (antennas in Mexico).
}
\end{figure}

\section{Results}
Fig. \ref{fig:uv_plane} presents examples of the \texttt {uv} distribution resulting from the MID and cropped MID configurations. We only display the cases for Dec values of +30 and --40 deg, to emphasize the clear differences between an ideal and a challenging case for the ngVLA. 
The relatively short on-source time helps to illustrate more clearly the consequences of removing the antennas in Mexico from the array; thus, short observations will be the most affected. Other ngVLA memos \citep[e.g.][]{rosero_65,jimenez_89a}, explore the properties of the PSF using longer integration times of $\sim$8 hours. 

As expected, there are important differences between ngVLA observations at northern declinations and those in the southern hemisphere. At Dec = +30 deg the {\texttt uv} distribution is close to circular, whereas at Dec = --40 deg it is considerably smaller in the North-South direction, resulting in a very elongated East-West distribution. This situation is exacerbated with the loss of antennas in Northern Mexico (green-colored data). Note that the antennas in Mexico also help to increase the density of data points at mid-length baselines, regardless of declination.

\subsection{PSF images}
\begin{figure}[!htb]
\centering
\includegraphics[width=1.0\textwidth]{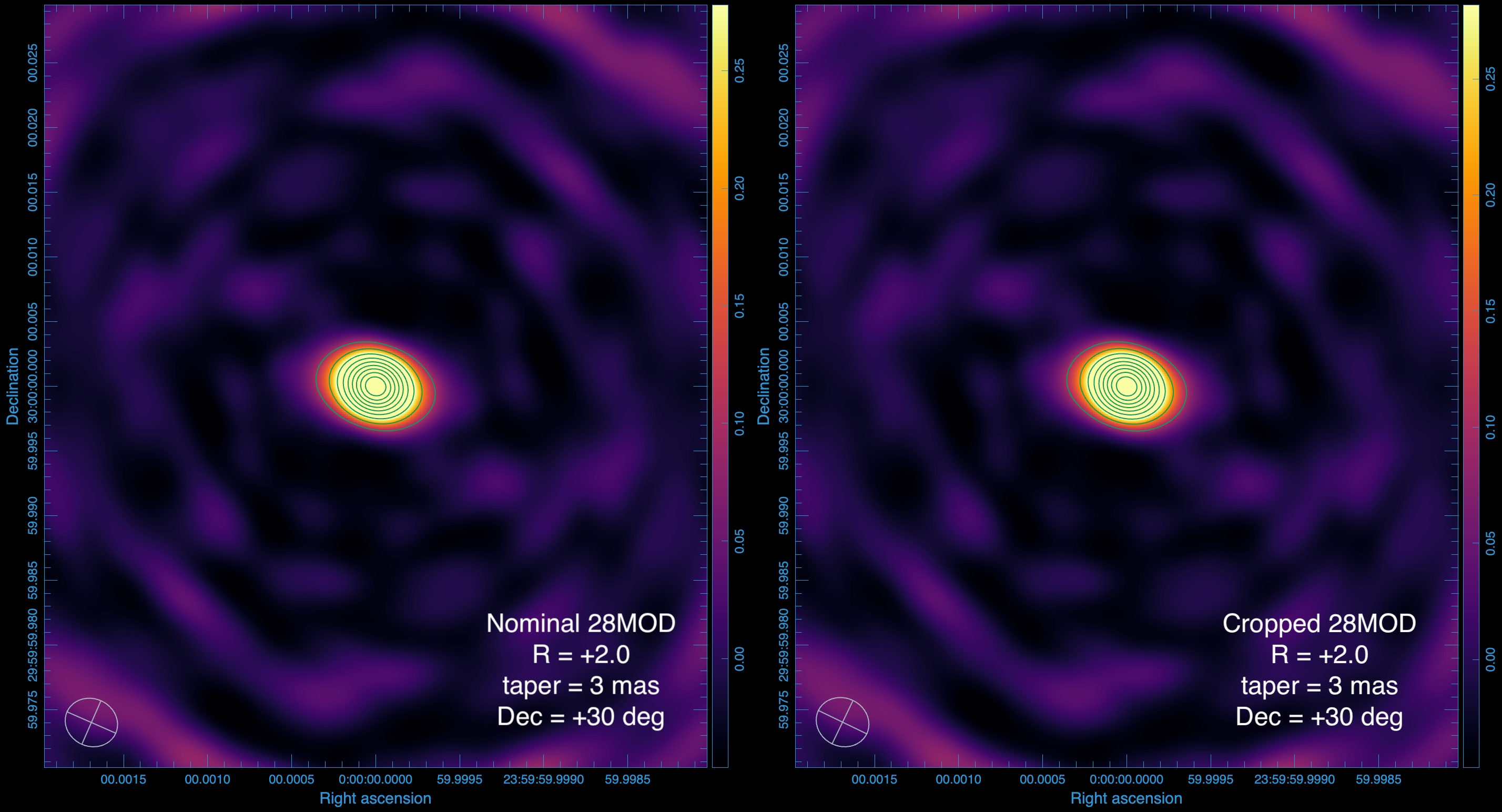}\\
\includegraphics[width=1.0\textwidth]{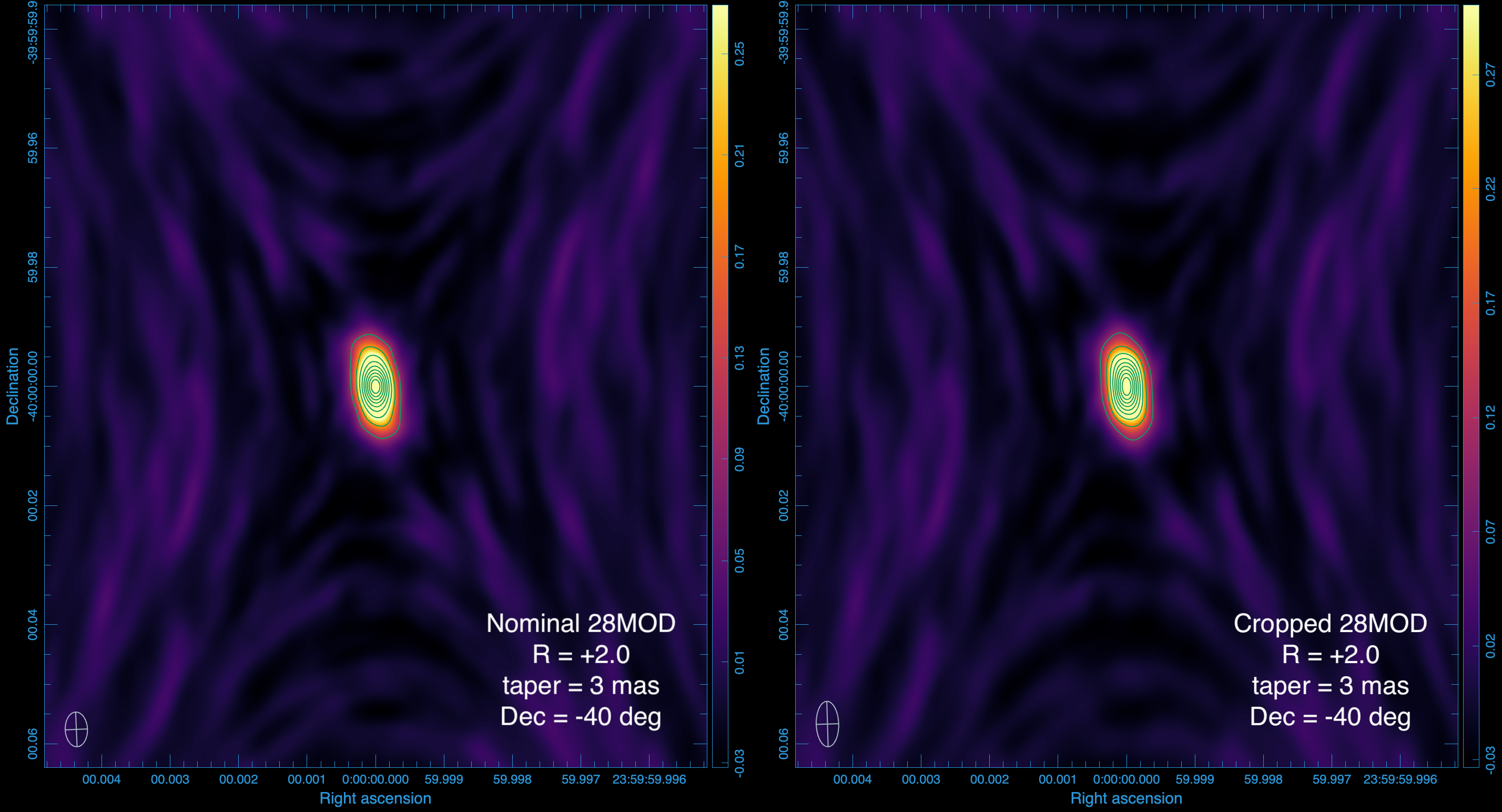}
\caption{\label{fig:psf_images_2} ngVLA Band 6 PSF images (R = +2.0; taper = 3.0 mas) resulting from the MID configuration with all antennas (\textit{left} panels) and without antennas in Mexico (\textit{right}). Cases with Dec = +30 deg (\textit{upper} panels) and Dec = --40 deg (\textit{lower}) are shown as representative cases. Contours are from 10 to 100\% of the peak value, in increments of 10\%. When observing in the southern hemisphere, removing the antennas in Mexico results in a larger ellipticity of the synthesized beam.}
\end{figure}

\begin{figure}[!htb]
\centering
\includegraphics[width=1.0\textwidth]{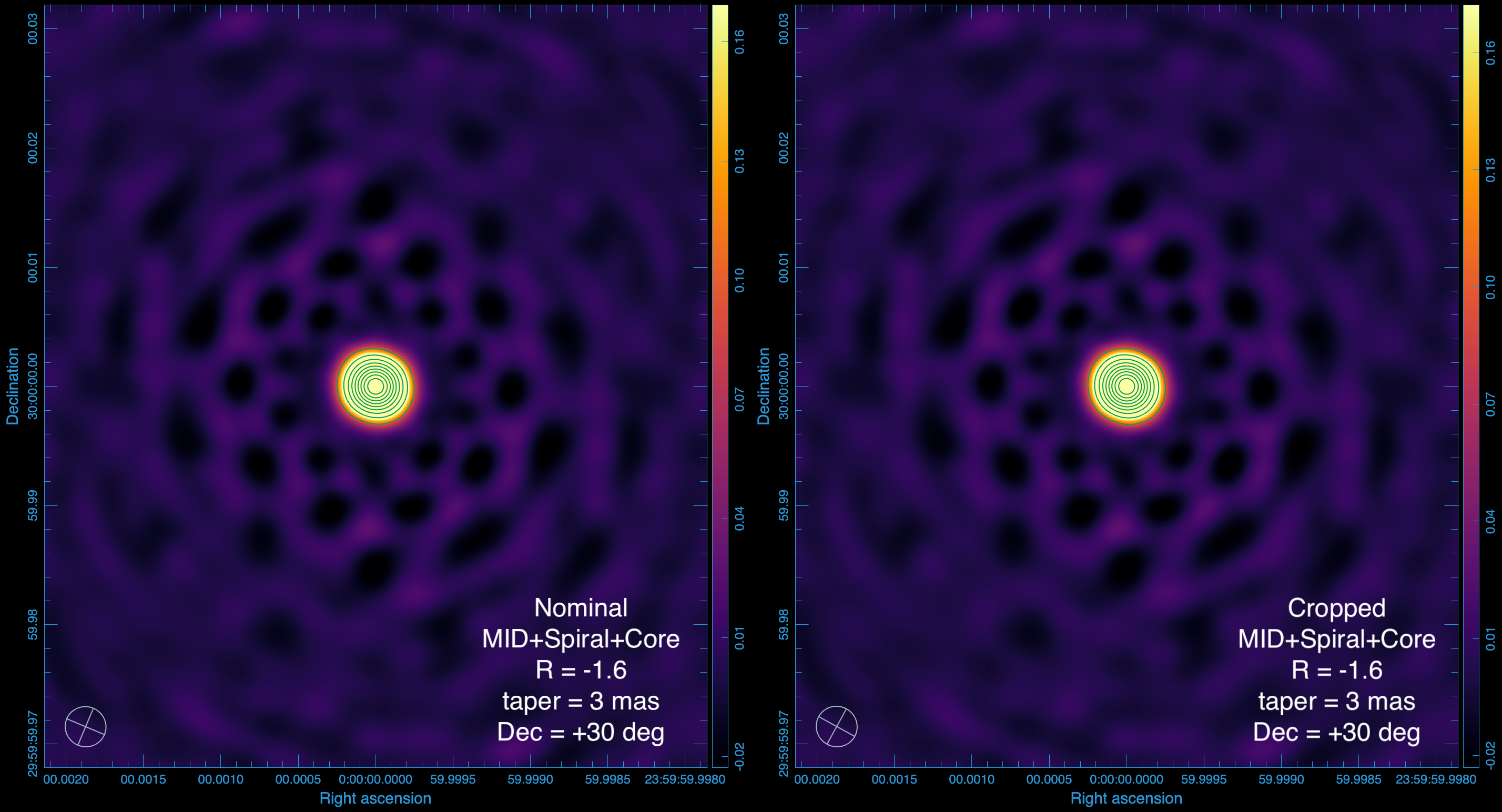}\\
\includegraphics[width=1.0\textwidth]{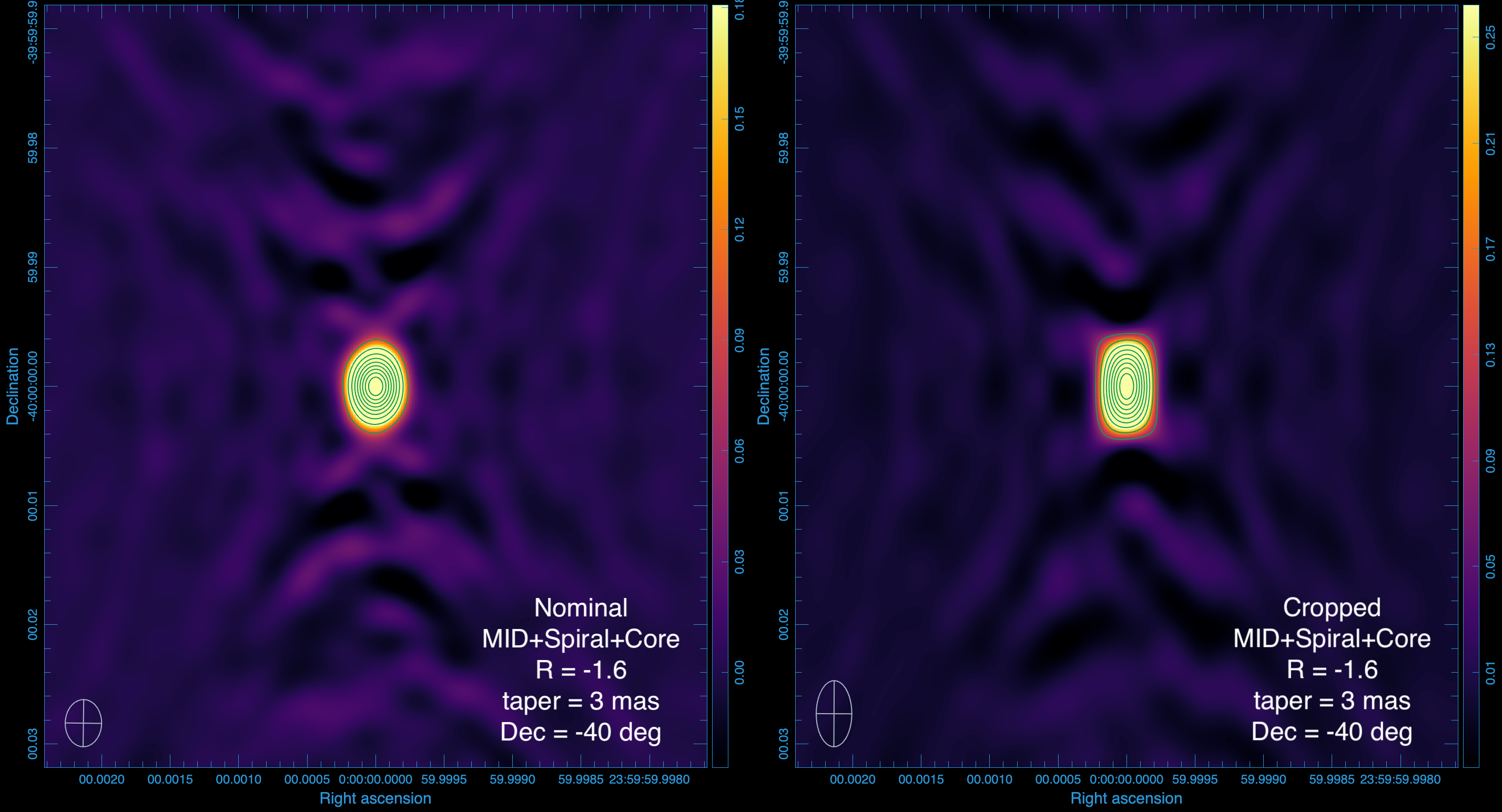}
\caption{\label{fig:psf_images_3} 
PSF images produced with MID+Spiral+Core (R = --1.6; taper = 3.0 mas), ordered in the same fashion as for Fig. \ref{fig:psf_images_2}. Contours are the also drawn from 10 to 100\% of the peak value, in increments of 10\%. As in the MID-only cases, we see similar results, where the absence of antennas in Mexico does not introduce any significant issues at a positive declination. In the southern hemisphere however, those antennas will significantly help to keep the PSF circularity and prevent artifacts around the main lobe.}
\end{figure}

For the same declinations shown in Fig. \ref{fig:uv_plane}, and by employing the MID configuration and its cropped version, the central regions of the PSF images (R = +2.0; taper = 3.0 mas) are displayed in Fig. \ref{fig:psf_images_2}. The \texttt{CARTA} visualization software \citep{CARTA_2021} was used with an arbitrary color scale per image to better visualize the morphology of the main lobe and the structure around it. 
In the northern hemisphere, as shown in the example at Dec = +30 deg (upper panels), the absence of the Mexican antennas will not pose a serious concern to the performance of the array. Image structure is similar between the two panels and without any obvious artefacts around the PSF core. At Dec = --40 deg (lower panels), the 10\% level contour occupies a larger area, signifying the more extended low-level emission outside the PSF core. While the structure around the main lobe is not completely different, the cropped configuration image has larger negative areas around the main lobe. It also produces a more elongated beam, which will in turn result in a larger geometric mean value. 
The situation is similar for the cases produced with the MID+Spiral+Core configurations (Fig. \ref{fig:psf_images_3}), using R = --1.6 and taper = 3.0 mas. Again, an extended 10\% level contour indicates that the PSF core starts to blend with its surroundings. All this confirms that not only a MID-only configuration will benefit by keeping the antennas in Mexico, as currently planned. In the next sections we compare the beam properties produced by nominal and cropped configurations. 

\subsection{PSF radial profiles}
In Fig. \ref{fig:psf_profiles_MID} and \ref{fig:psf_profiles_MIDplus} we present examples of the PSF radial profiles for MID (R = +2.0; taper = 3.0 mas) and MID+Spiral+Core (R = --1.6; taper = 3.0 mas), as well as for the cropped versions. The profiles are azimuthally averaged and produced with the Python package {\texttt{photutils} for aperture photometry. 
In the figures, we compare the PSF profiles with the corresponding gaussian response built using the {\texttt{tclean} fitting (clean beam) results, i.e.  using the geometric mean of the beam semi major and semi minor axes as the gaussian HWHM. 
In this memo, we mainly look at how the PSF gets degraded by removing the antennas in Mexico, so the images are compared one-to-one, i.e. produced by the same robust and tapering parameters. 

At Dec = +30 deg, the gaussian fitting parameters from \texttt {tclean} produce a (clean beam) profile that follows well the actual PSF profile for both MID configurations (Fig. \ref{fig:psf_profiles_MID}, left panels). Both HWHM values and the subtraction of the PSF profiles are within 1\%.
However, at Dec = --40 deg (right panels) the difference between the two MID configurations is  obvious. The beam HWHM is now $\sim$15\% larger when removing the antennas from Mexico, and the subtraction of the PSF profiles peaks at around $\sim$5\%.
A similar behavior is seen when comparing the MID+Spiral+Core configurations (Fig. \ref{fig:psf_profiles_MIDplus}). 
The differences in terms of beam HWHM and PSF profile subtractions are at very similar levels as for the MID-only case at Dec = --40 deg, $\sim$17\% and $\sim$7\%, respectively. Again, at Dec = +30 deg there is no significant difference between the two PSF profiles.
In the next section we look at the ellipticity, where a more obvious trend in beam properties can be seen.

\begin{figure}[!htb]
\centering
\includegraphics[width=0.50\textwidth]{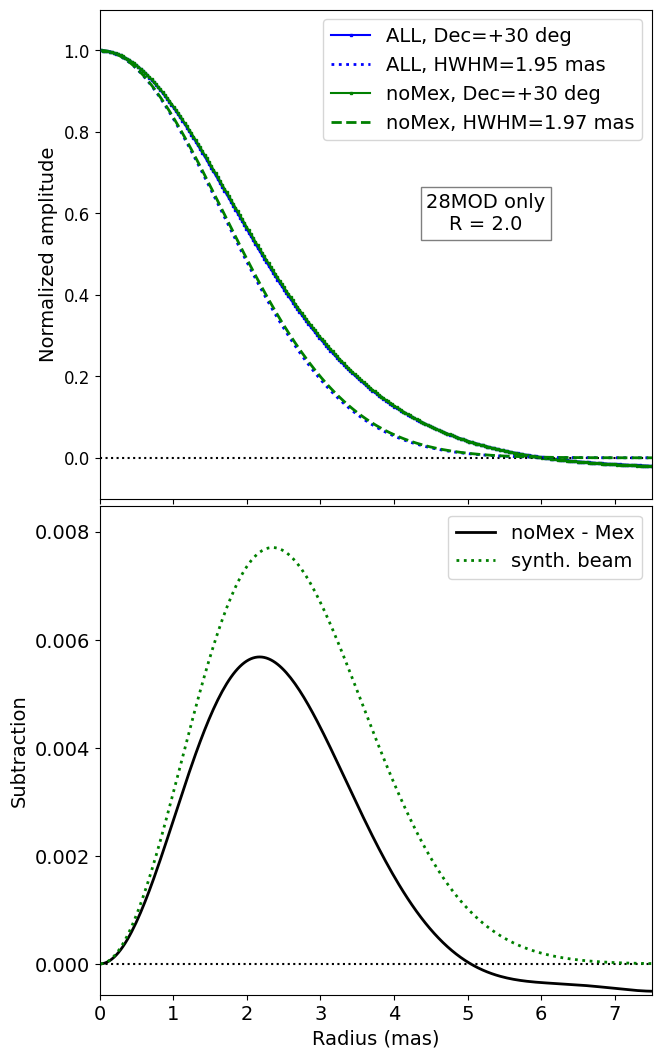}
\includegraphics[width=0.49\textwidth]{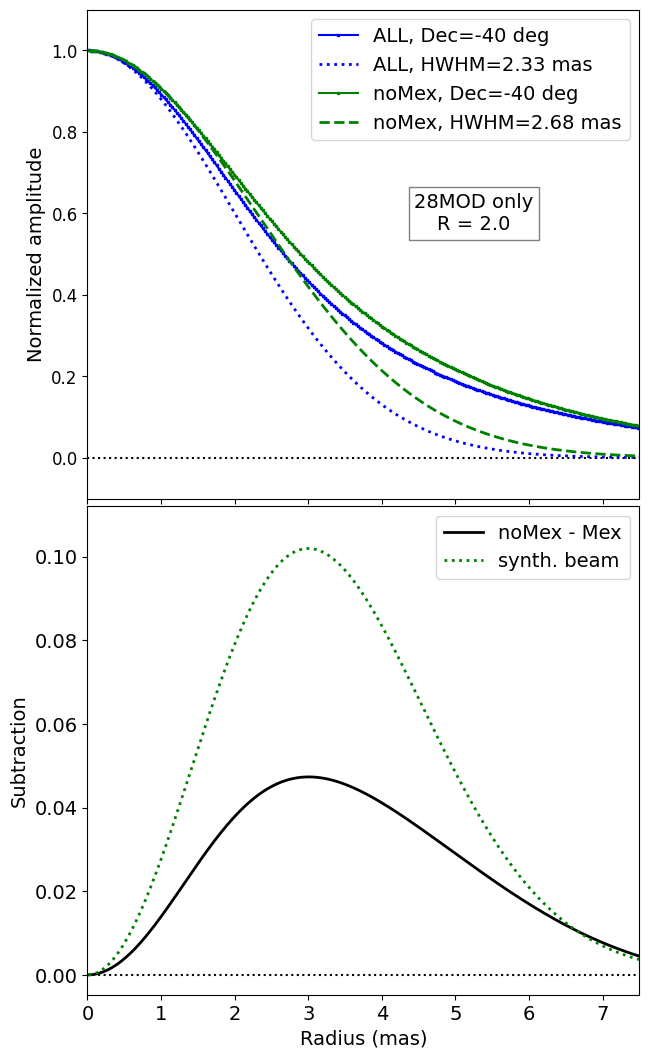}
\caption{\label{fig:psf_profiles_MID} 
\textit{Up:} Azimuthally averaged radial profiles of the ngVLA PSF (R = +2.0; taper = 3.0 mas) for Band 6. Blue data points present cases with all antennas in the MID configuration and green those from cropped versions. The \texttt {tclean} fitted beams, with  HWHM values indicated, are shown in dashed and dotted lines. Cases shown are for Dec = +30 (\textit{left}) and --40 deg (\textit{right}) and produced with the images in Fig. \ref{fig:psf_images_2}. \textit{Bottom:} Subtraction between corresponding beams and PSF profiles. For these particular cases, at Dec = --40 deg, the PSF subtraction peaks at $\sim$5\% at $\sim$ 3 mas and the HWHM is $\sim$15\% larger when removing antennas from Northern Mexico. At Dec = +30 deg, there is no significant difference.}
\end{figure}

\begin{figure}[!htb]
\centering
\includegraphics[width=0.5\textwidth]{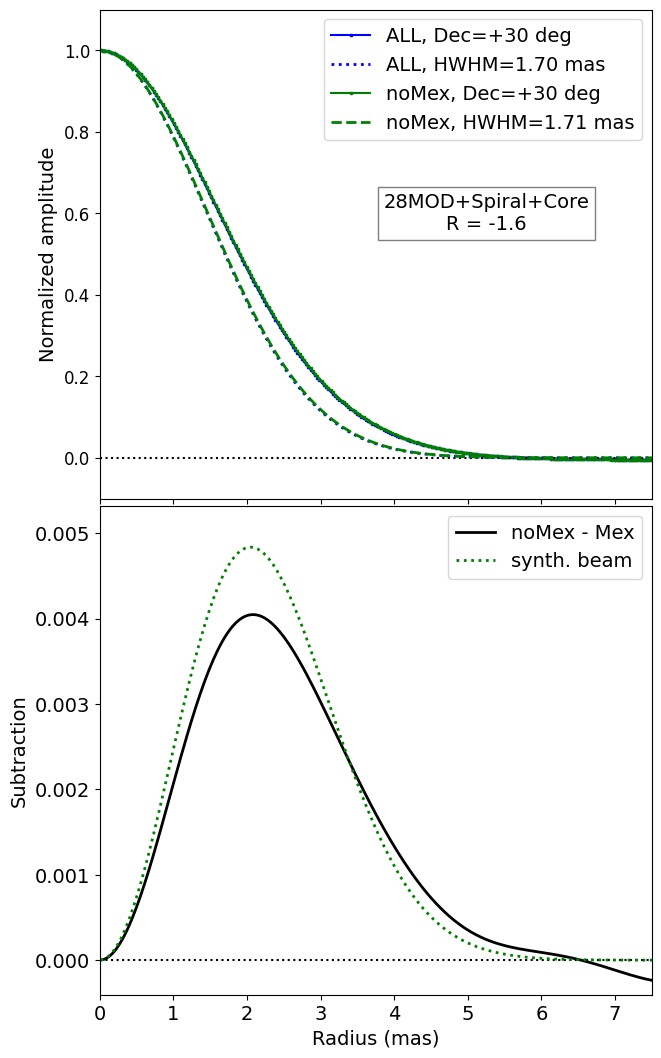}
\includegraphics[width=0.49\textwidth]{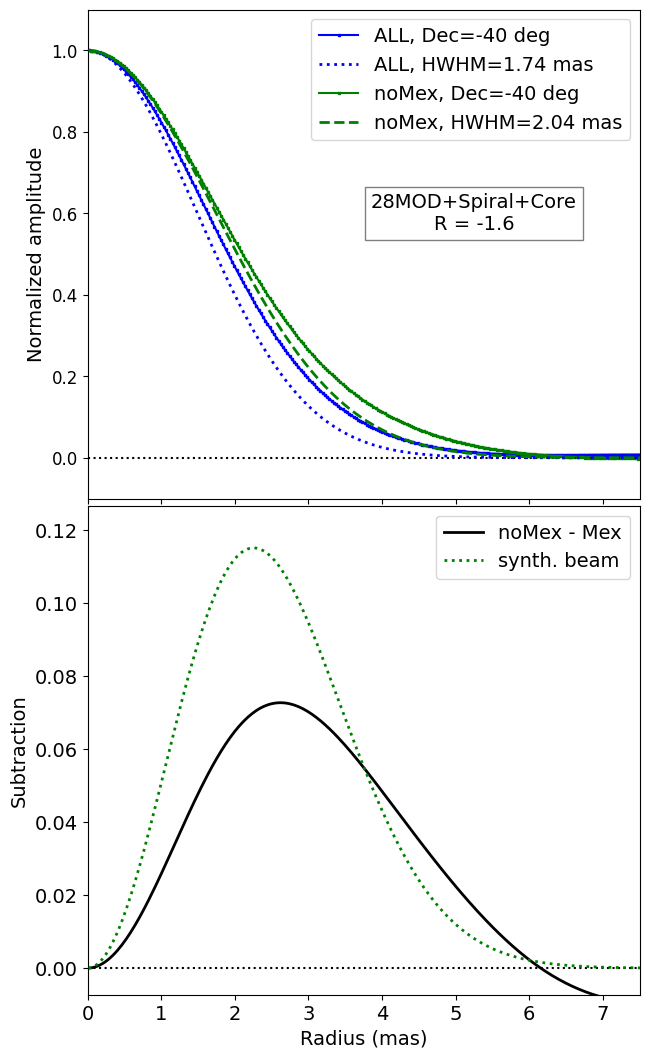}
\caption{\label{fig:psf_profiles_MIDplus} Corresponding results for the MID+Spiral+Core cases (R = --1.6; taper = 3.0 mas), produced with the PSF images in Fig. \ref{fig:psf_images_3}. The trends in HWHM values and the subtraction of the PSF profiles largely follow the results for the MID-only cases. At Dec = --40 deg (\textit{right}), the beam is 17\% wider and the subtraction peaks at $\sim$7\%. At Dec = +30 deg (\textit{left}), the two configurations have very similar PSF and beams and their subtraction stays within 1\%.}
\end{figure}

\subsection{PSF ellipticity}
\label{sec:ellip}
We are in particular interested in looking at the performance of the array for cases at negative declinations, for both MID and MID+Spiral+Core. Cases with smaller beam ellipticity are of course those at positive declinations. With this, and the fact that the ellipticity has a more clear dependence on the robust values, and to a less degree on the configuration used, we look into more detail on the robust and declination parameter space.

Fig. \ref{fig:psf_ellip} presents (taper = 3.0 mas), as a function of Declination, the beam ellipticity and its ratio between the cropped and nominal configurations. 
For the MID cases (upper panels), and for all values in the range Dec $<$ 0 deg, cropped-version images have a larger beam ellipticity, as compared to those of the nominal configuration. For example, at Dec = --40 deg it is $\sim$ 2.0 vs $\sim$ 1.6, for all robust values used (1.2 to 2.0). 
The mean values (dotted and dashed lines) provide a confirmation of a larger ellipticity trend for the cropped configurations, especially at southern declinations. 
As expected, at positive declinations the beam ellipticity is small at $\sim$ 10\% above the unity for both configurations.
To see more clearly the change between cropped and nominal configurations, we look at their ellipticity ratio (right panel). It is clear that the beam will degrade (ellipticity wise) by as much as $\sim$ 30\% for Dec = --40 deg; at Dec = --20 this figure decreases to $\sim$ 7\%. At positive declinations, the ratio essentially becomes unity.

MID+Spiral+Core is a very different configuration than MID with its large number of antennas at short baselines. We note of an increment (Fig. \ref{fig:psf_ellip} lower left panel) in the mean ellipticity, for both cropped and nominal versions, of $\sim$10-20\% compared to the MID-only case, depending on Declination. Thus, compared to MID, the beam ellipticity for MID+Spiral+Core is more affected for the observations/imaging setup used. The imaging results in similar ellipticity ratios (lower right panel) compared MID, per Declination. We find a variation of $\sim$ 5\% between the two, based on the mean values. 

Finally, Fig. \ref{fig:psf_ellip_2} shows corresponding results for the tests with a taper of 4.0 mas. The same trends, for cropped configurations, of higher ellipticity and its ratio at southern declinations are found. However, all values are smaller overall when compared to the 3.0 mas taper setting, e.g., with 4.0 mas the most extreme cases have ellipticities of $\sim$ 1.8 and 2.0 (vs $\sim$ 2.0 and 2.2 for taper = 3.0 mas) at Dec = --40 deg, for cropped versions of MID and MID+Spiral+Core, respectively. The same trend is observed when looking at the mean ellipticity. In terms of the ellipticity ratio (right panels), we find values in the range of $\sim$ 1.2-1.3, for Dec = --40 deg, depending on the robust value and configuration. 
Overall, the synthesized beams obtained with a taper of 3.0 mas degrade (ellipticity wise) in average $\sim$ 10\% (at Dec $\leq$ --20 deg) when removing the antennas from Mexico, as compared to a taper of 4.0 mas. On the side of positive declinations, the difference in ellipticity ratio between tapering values is not significant. We can see that the mean (dotted lines) basically aligns with the unity (vertical dashed line). Table \ref{tab:ellipticity} presents mean values for the synthesized beam major and minor axis and its ellipticity for --25 $\leq$ Dec $\leq$ --40 deg.

We tested the potential impact of the image size used to properly grid all visibilities are the shortest baselines. Additional synthetic observations were obtained using a pixel and map sizes of 0.054 mas and 16384 pixels (or 885 mas). This test provided maps four times larger in angular units. While the \texttt{CASA} fitted beams are not exactly the same as for the cases presented and discussed in this memo (0.027 mas/8192 pixels), the general trend of ellipticity and its ratio holds, i.e. the cropped configurations consistently show larger ellipticities at negative declination values.

\begin{table}[!th]
	\centering
	\caption{A sample of mean values for the synthesized beam parameters (taper of 3 mas) for the nominal and cropped versions of MID and MID+Spiral+Core.}
	\label{tab:ellipticity}
	\begin{tabular}{ccccccc}
    	\hline\hline
        Dec & Major & Minor &  Ellipticity & Major & Minor&  Ellipticity \\
		 (deg)  &    (mas)     & (mas) & &    (mas)     & (mas) &      \\
    	\hline
        & \multicolumn{3}{c}{MID} & \multicolumn{3}{c}{Cropped MID}\\
        \hline
        -25.0 & 4.6 & 3.8 & 1.2 & 5.0 & 3.8 & 1.3\\
        -30.0 & 4.8 & 3.8 & 1.3 & 5.5 & 3.8 & 1.4\\
        -35.0 & 5.2 & 3.7 & 1.4  & 6.3 & 3.8 & 1.7\\
        -40.0 & 5.7 & 3.7 & 1.6 & 7.5 & 3.7 & 2.0\\
        \hline
        & \multicolumn{3}{c}{MID+Spiral+Core} & \multicolumn{3}{c}{Cropped MID+Spiral+Core}\\
        \hline
        -25.0 & 4.6 & 3.7 & 1.2 & 5.0 & 3.7 & 1.4\\
        -30.0 & 4.7 & 3.6 & 1.3 & 5.4 & 3.6 & 1.5\\
        -35.0 & 4.9 & 3.5 & 1.4 & 6.0 & 3.5 & 1.7\\
        -40.0 & 5.2 & 3.4 & 1.5 & 7.0 & 3.4 & 2.0\\
        \hline
\end{tabular}
\end{table}
{\footnotesize All mean values are calculated using all robust values, and therefore correspond to the dash-dotted lines in the left panels of Figs. \ref{fig:psf_ellip} and \ref{fig:psf_ellip_2}. Only one decimal is used for simplicity.}
\begin{figure}[!htb]
\centering
\includegraphics[width=1.0\textwidth]{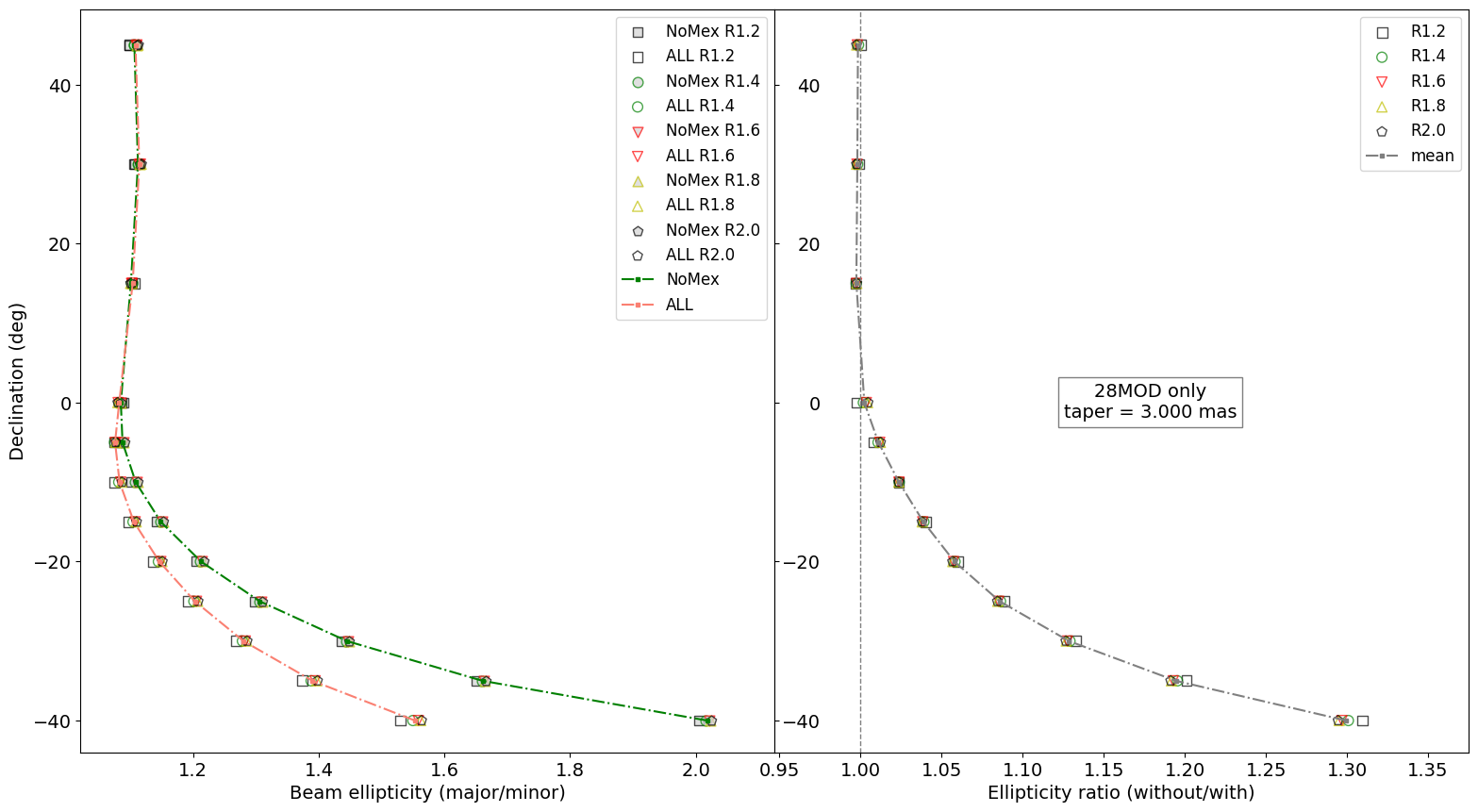}\\
\includegraphics[width=1.0\textwidth]{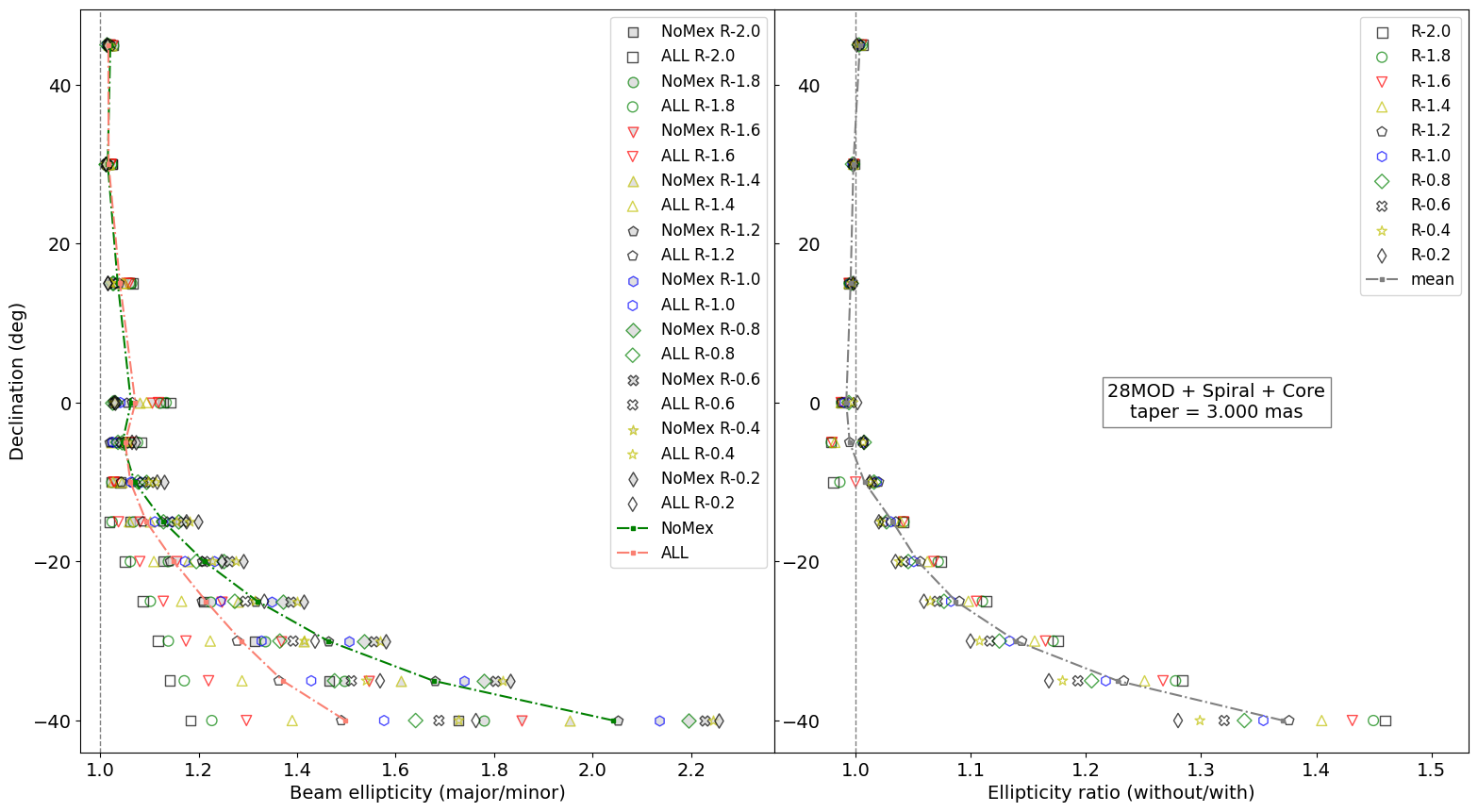}
\caption{\label{fig:psf_ellip} Band 6 beam ellipticity (\textit{left}) and the ratio of it for the cropped configuration over the nominal one (without/with; \textit{right}), for all robust values. Panels present cases for taper = 3.0 mas using the MID (\textit{upper}) and 28MOD+Spiral+Core (\textit{lower panels}) configurations, respectively. Mean values for nominal and cropped configurations are shown in dash-dotted lines.
Cropped configurations at the majority of robust cases show a very clear trend of significantly larger ellipticity at southern declinations. For the nominal configurations, the less affected cases have the more negative robust values. The ellipticity ratio shows that all observations in the southern hemisphere, e.g. Dec $\leq\sim$ --20 deg, will benefit when including all the antennas in Mexico. Note the slight fluctuations in the ellipticity ratio around the unity for 0$\leq$ Dec $\leq$--10 deg for the most negative robust values.
}
\end{figure}

\begin{figure}[!htb]
\centering
\includegraphics[width=1.0\textwidth]{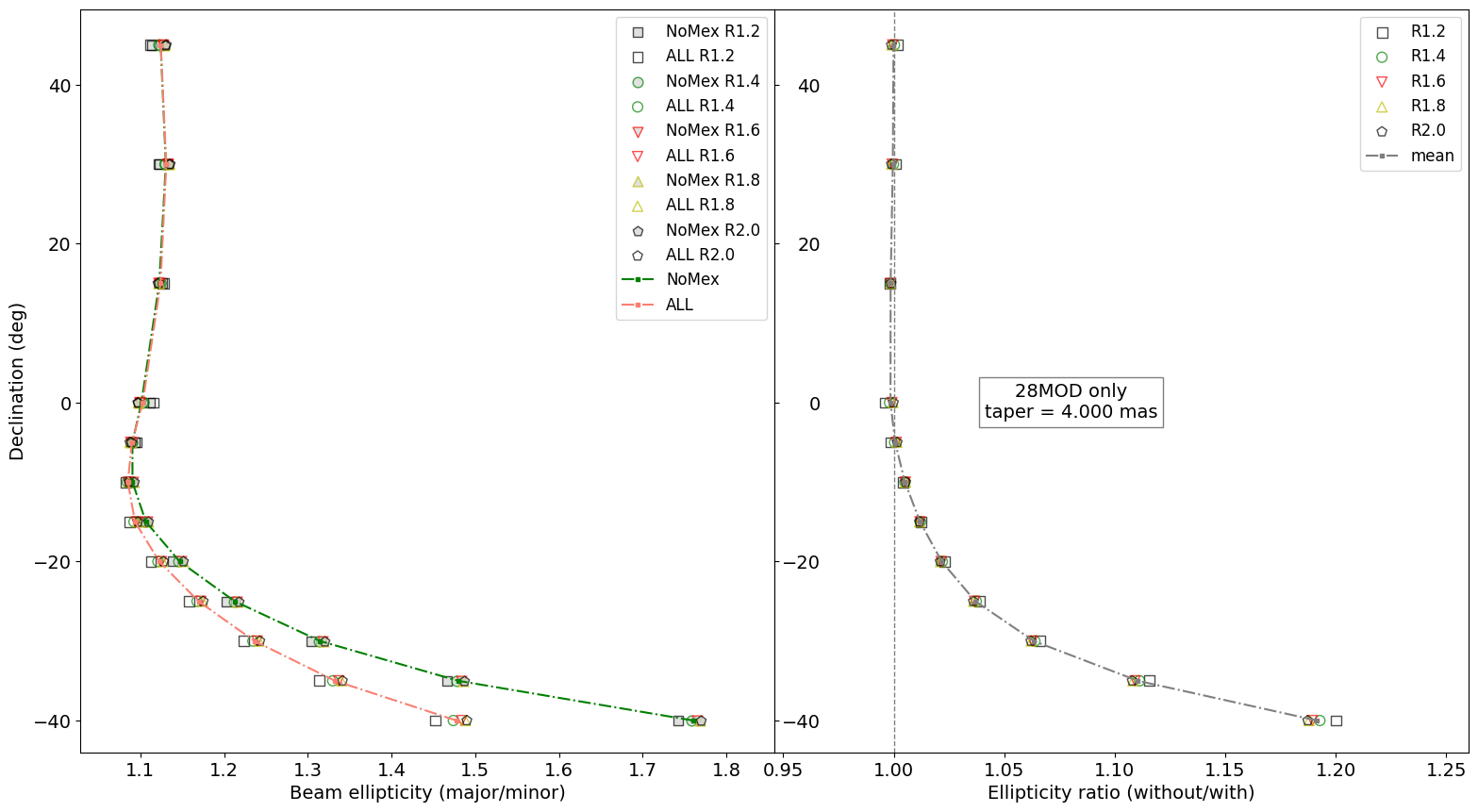}\\
\includegraphics[width=1.0\textwidth]{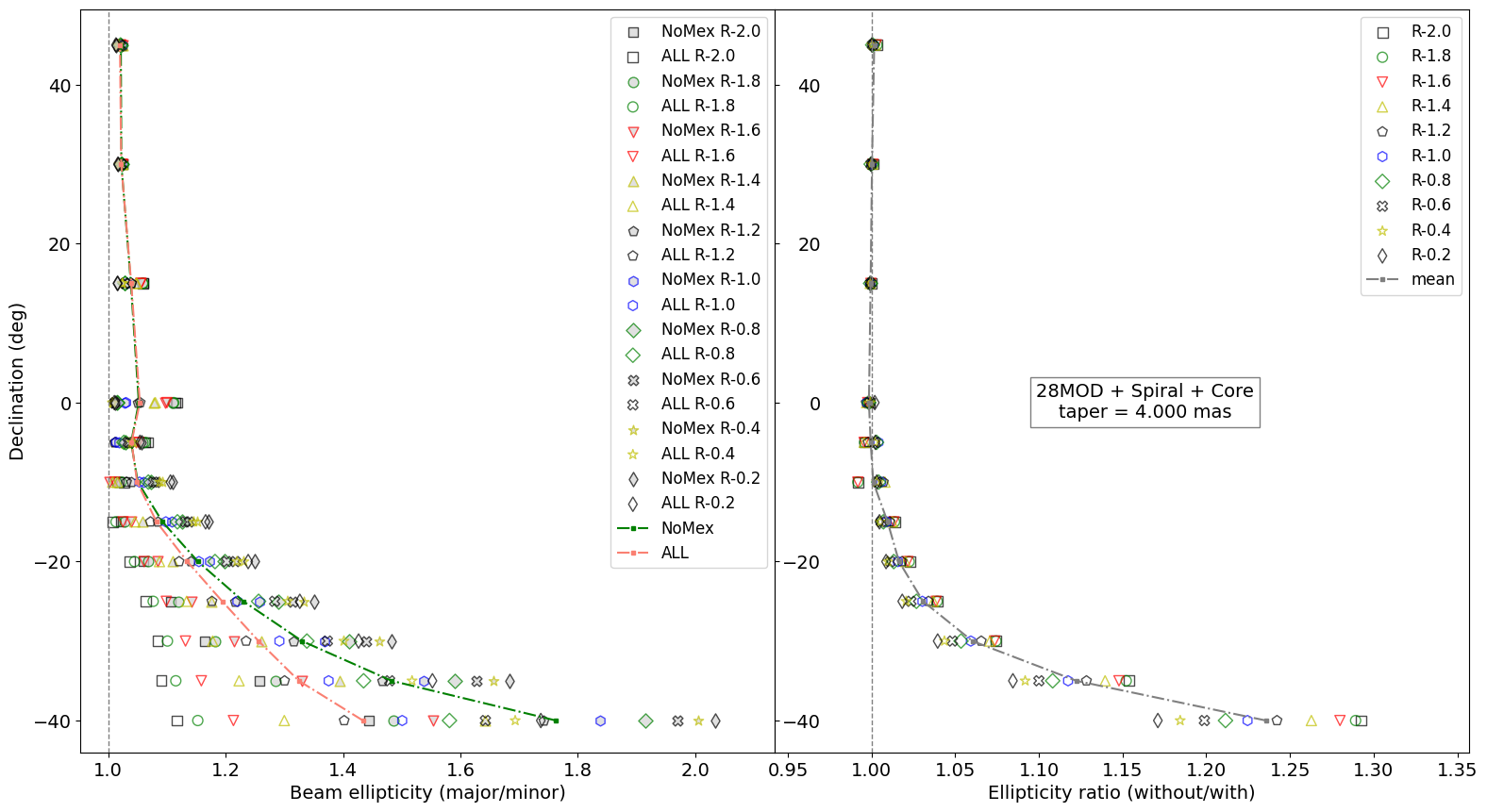}
\caption{\label{fig:psf_ellip_2} Similar as in Fig. \ref{fig:psf_ellip}, we display results for the cases of taper = 4.0 mas. With this larger taper value, a slightly larger and less elliptical beam is obtained, helping to decrease the ellipticity ratio (right panels). 
As expected, more positive robust values produce a larger beam, decreasing the ellipticity and the difference of that between nominal and cropped configurations. However, observations in the southern hemisphere continue to be benefited by the inclusion of all antennas in Mexico. Note that with this tapering, the ellipticity fluctuations at 0$\leq$ Dec $\leq$--10 deg are largely gone. }
\end{figure}

\section{Conclusions}
Removing antennas from an interferometric array usually affects the PSF image quality. In this memo we showed that by removing the antennas in Mexico (with the exception of T27 which falls very close to the Mexico--U.S.A. border), several PSF properties worsen considerably at Dec $\lesssim$ --20 deg. In general, median values of the ellipticity increase, for all tapering values used. With respect to the nominal configurations and for a tapering of 3.0 mas, we see mean ellipticity (over the robust space) increments of $\sim$ 1.3$\times$ and 1.4$\times$ for a declination of --40, for the MID and MID+Spiral+Core cases, respectively. At 4.0 mas, the above values become $\sim$ 1.2$\times$ and 1.3$\times$.
At positive declinations there is essentially a negligible penalty for the exclusion of antennas in Mexico.

Although larger tapering values will help to reduce the ellipticity for cropped configurations at all declinations, it remains less effective at the southern hemisphere and will impose more constraints in terms of angular resolution.
Thus, higher resolution scientific cases, targeting sources in the southern hemisphere, will be the most affected if the most southern antennas (those in Northern Mexico) of the ngVLA are not deployed.

\section*{Acknowledgements}
We thank Viviana Rosero, Chris Carilli, and Eric Murphy (NRAO) for very useful discussion and feedback that helped to improve the content and reach of this work.

\bibliographystyle{apj_w_etal_3auth}
\bibliography{sample}

\end{document}